\PassOptionsToPackage{table}{xcolor}
\documentclass[sigconf]{acmart}

\usepackage{booktabs}
\usepackage[utf8]{inputenc}
\usepackage{multirow}
\usepackage{graphicx}
\usepackage{xspace}
\usepackage{pifont}

\usepackage{nicematrix}
\usepackage{tcolorbox}

\definecolor{purpleheart}{rgb}{0.41, 0.21, 0.61}
\newif\ifshowcomment
\showcommentfalse 
\ifshowcomment
\newcommand{\jce}[1]{\textcolor{purpleheart}{[JC: #1]}}
\newcommand{\todo}[1]{{\color{red} {[TODO: #1]}}}
\newcommand{\update}[1]{{\color{blue} {#1}}}
\newcommand{\tara}[1]{{\color{violet} {[Tara: #1]}}}

\newcommand{\kurt}[1]{{\color{teal} {[Kurt: #1]}}}
\newcommand{\pgk}[1]{{\color{brown} {[PGK: #1]}}}
\newcommand{\renee}[1]{{\color{brown} {[RS: #1]}}}
\newcommand{\ar}[1]{{\color{cyan} {[AR: #1]}}}
\newcommand{\sarah}[1]{{\color{purple} {[Sarah: #1]}}}

\else
\newcommand{\jce}[1]{}
\newcommand{\todo}[1]{}
\newcommand{\update}[1]{}
\newcommand{\tara}[1]{}
\newcommand{\amw}[1]{}
\newcommand{\kurt}[1]{}
\newcommand{\pgk}[1]{}
\newcommand{\renee}[1]{}
\newcommand{\ar}[1]{}
\newcommand{\sarah}[1]{}
\fi

\newcommand{\dataASize}{637,600 }
\newcommand{\dataBSize}{13,800 }

\copyrightyear{2026}
\acmYear{2026}
\setcopyright{cc}
\setcctype{by}
\acmConference[CHI '26]{Proceedings of the 2026 CHI Conference on Human Factors in Computing Systems}{April 13--17, 2026}{Barcelona, Spain}
\acmBooktitle{Proceedings of the 2026 CHI Conference on Human Factors in Computing Systems (CHI '26), April 13--17, 2026, Barcelona, Spain}
\acmPrice{}
\acmDOI{10.1145/3772318.3790556}
\acmISBN{979-8-4007-2278-3/2026/04}

\newcommand{\pquote}[2]{%
  \begin{quote}
    \textit{``#1''}%
    \ifthenelse{\equal{#2}{}}{}{~(#2)}%
  \end{quote}}

\newcommand{\inlinequote}[1]{\textit{``#1''}}
\newcommand{\eg}{e.g., }

\newcommand{\ie}{i.e., }
\newcommand{\etal}{et al.\@\xspace}
\newcommand{\subreddit}[1]{\emph{#1}\xspace}

\definecolor{lightgray}{gray}{0.95}
\definecolor{customtaupe}{HTML}{aecbfa}

\begin{document}

\title[Understanding People's Emotions and Help Needs During Scams]{``It didn't feel right but I needed a job so desperately'':\\ Understanding People's Emotions \& Help Needs \\During Financial Scams}

\author{Jake Chanenson}
\orcid{0000-0003-3400-925X}
\affiliation{%
  \institution{Google}
  \country{USA}
}
\additionalaffiliation{%
  \institution{University of Chicago}
}
\email{jchanen1@uchicago.edu}

\author{Tara Matthews}
\orcid{0009-0008-4805-8339}
\affiliation{%
  \institution{Google}
  \country{USA}
}
\email{taramatthews@google.com}

\author{Sunny Consolvo}
\orcid{0000-0001-5337-328X}
\affiliation{%
  \institution{Google}
  \country{USA}
}
\email{sconsolvo@google.com}

\author{Patrick Gage Kelley}
\orcid{0000-0003-4405-0010}
\affiliation{%
  \institution{Google}
  \country{USA}
}
\email{patrickgage@acm.org}

\author{Jessica McClearn}
\orcid{0000-0002-1609-5257}
\affiliation{%
  \institution{Google}
  \country{USA}
}
\additionalaffiliation{%
  \institution{Royal Holloway, University of London}
}
\email{jessica.mcclearn.2021@live.rhul.ac.uk}

\author{Sarah Meiklejohn}
\orcid{0000-0001-5671-1395}
\affiliation{%
  \institution{Google}
  \country{USA}
}
\additionalaffiliation{%
  \institution{University College London}
}
\email{s.meiklejohn@ucl.ac.uk }

\author{Abhishek Roy}
\orcid{0009-0001-5865-8740}
\affiliation{%
  \institution{Google}
  \country{USA}
}
\email{abhishekroy@google.com}

\author{Renee Shelby}
\orcid{0000-0003-4720-3844}
\affiliation{%
  \institution{Google}
  \country{USA}
}
\email{reneeshelby@google.com}

\author{Kurt Thomas}
\orcid{0000-0002-3762-5851}
\affiliation{%
  \institution{Google}
  \country{USA}
}
\email{kurtthomas@google.com}

\author{Amelia Hassoun}
\orcid{0000-0002-6328-4763}
\affiliation{%
  \institution{Google}
  \country{UK}
}
\additionalaffiliation{%
  \institution{University of Cambridge}
}

\email{miahassoun@google.com}

\renewcommand{\shortauthors}{Chanenson et al.}



\begin{abstract}
    Online financial scams represent a long-standing and serious threat for which people seek help. We present a study to understand people’s in situ motivations for engaging with scams and the help needs they express before, during, and after encountering a scam. We identify the main emotions scammers exploited (e.g., fear, hope) and characterize how they did so. We examine factors---such as financial insecurity and legal precarity---which elevate people’s risk of engaging with specific scams and experiencing harm. We indicate when people sought help and describe their help-seeking needs and emotions at different stages of the scam. We discuss how these needs could be met through the design of contextually-specific prevention, diagnostic, mitigation, and recovery interventions.
\end{abstract}

\maketitle

\section{Introduction}
Online financial \emph{scams}\footnote{Throughout this paper, we use \textit{scams} to refer to \textit{online financial scams}, which is synonymous with \textit{online financial fraud}, unless otherwise specified.} represent a long-standing and serious threat \cite{aarp2025, sheng2010, norris2019, titus2001, breen2022, button2017, button2009,coluccia2020}. Losses reported to the U.S. Federal Trade Commission (FTC)  exceeded \$12.5 billion USD in 2024 ~\cite{ftc2025}. However, truly accounting for the impact of scams is challenging, as most incidents go unreported and many harms are not financial~\cite{houtti2024, anderson2021}. Scammers' tactics are constantly evolving and growing in sophistication~\cite{zhu2021, breen2022} as they attempt to engage and exploit targets\footnote{We use the term \emph{targets} to refer to people a scammer attempts to engage, engages, or harms. We avoid the term \textit{victims}, because it has been critiqued as being associated with negative attributes, such as powerlessness or weakness, that can be both stigmatizing and internalized~\cite{oshea2024victim}.}~\cite{oak2025pig, houtti2024, button2017, button2009, coluccia2020, hanoch2021}. Targets span all ages, genders, and demographics~\cite{breen2022, houtti2024, ftc2025, balakrishnan2025, titus2001, sheng2010}. Understanding why and how people fall prey to scams, and what help they need in critical moments before, during, and after encountering a scam, is necessary to mitigate this substantial harm.

Existing HCI and computer security research has investigated specific scam types---including romance~\cite{amirkhani2024, carter2021, cross2023, cross2018, whitty2016}, pig butchering~\cite{oak2025pig}, and e-commerce~\cite{blancaflor2023, bitaab2023}, among others~\cite{beals2015}---but relatively few studies have examined the experiences and needs of targets across a range of scam types and scammer tactics as scams unfold~\cite{oak2025reddit,bouma2025ccs}. Emerging research details how online communities provide emotional support and vetted guidance toward scam recovery, also identifying knowledge deficits that increase  a target's susceptibility~\cite{oak2025reddit,bouma2025ccs}. Building on this work, we focus on targets’ emotional states at key moments during a scam---such as when they are lured in (or ``hooked'' by the scammer), realize something is wrong, or ask for help. Our goal is to understand why targets engage with specific scams, why some targets experience elevated risk, and what types of help targets need at different stages of the scam to avoid, diagnose, mitigate, and recover from harm.

We examine 405 posts from Reddit that involve seeking help for a range of known and emerging scams. Posters commonly describe their encounters, providing rich descriptions of their emotions and needs at different stages of the scam. Our analysis builds on existing frameworks and research~\cite{matthews2025, wei2024, beals2015, scheuerman2021, roy2025} to provide practical empirical insights. Systematically examining a range of scam types helps identify patterns in what motivates targets to engage and how scammers manipulate those motivations. To inform the design and development of effective scam interventions, we investigated the following research questions:

\begin{itemize}
    \item [RQ1.] What \emph{motivates} people to consider engaging with scams and ultimately to comply with scammer requests?
    \item [RQ2.] \emph{When} during a scam do people seek help? What are their \emph{emotions and help needs} at different stages of the scam?
    \item [RQ3.] How do \emph{contextual risk factors} augment scam susceptibility and harm?
    \item [RQ4.] How might people's emotions and needs at different stages of the scam be used to \emph{inform interventions}?
\end{itemize}

Our primary contribution is a qualitative analysis of the emotions and help needs people expressed while experiencing scams, especially the emotional motivations they had for engaging with scams. We found five main \emph{emotional motivations} across 12 types of scams that scammers manipulated to engage targets: Fear, Guilt \& Goodness, Trust, Hope, and Belonging. Understanding these underlying motivations (like the hope for a job in a difficult economy) helps explain why certain people (such as those who are financially insecure) experience an elevated risk of being targeted, engaging with, and harmed by~\cite{warford2022} specific scams. 

We also propose an incremental expansion of the User States Framework~\cite{matthews2025}\footnote{The User States Framework~\cite{matthews2025} maps a user’s expected needs and emotional state of mind before, during, and after technology-facilitated attacks into the states of Prevention, Monitoring, Active Event, and Recovery.} for scams, adding nuance to the Active Event state. We use this framework and build on prior help-seeking research~\cite{wei2024} to map users' help-seeking needs---Sensemaking, Guidance, Therapeutic, and External Action---to their emotional context and stage in the scam's lifecycle. 

Our method enables us to contribute novel findings on targets' emotions and needs in the midst of an active scam. Overall, this work identifies opportunities for in situ scam interventions to help people effectively prevent, diagnose, mitigate, and recover from scams.
\section{Related Work}\label{sec:related}

We situate our work in the broader context of scam frameworks, the motivations known to affect how and why targets engage with scams, and online help-seeking patterns for scams.

\subsection{Scam Frameworks}

Previous research has produced scam frameworks reflecting different organizing principles. Some categorize scams by their structural or technical elements. Beals et al. \cite{beals2015} developed a five-level taxonomy that, at its highest level, distinguishes between fraud targeting individuals and organizations. Their sub-categories for individual fraud include consumer investment fraud, consumer products and services fraud, employment fraud, prize and grant fraud, phantom debt collection fraud, charity fraud, and relationship and trust fraud. Similarly, other frameworks classify scams by the channel through which they occur (e.g., web, mobile, telephone, physical)~\cite{onwubiko2020fraud}, the specific actions and deceptions involved in their execution (e.g., visceral appeals or pressure and coercion)~\cite{button2009}, or the primary narrative used for persuasion, such as a threat, an offer, or a promise~\cite{deliema2023a}.
This structural understanding of scams is strengthened by research on human factors, with Levi \cite{levi2008} analyzing the fluid, networked organization of perpetrators, and Button et al. \cite{button2009} synthesizing research on the characteristics and vulnerabilities of targets. Together, existing scholarship demonstrates that a comprehensive understanding of scams must integrate analyses of scam mechanics (including scammer tactics) and target experiences (including their motivations, differentiated risks, and support needs).

Scam impacts can extend beyond financial harm to inflict emotional, relational, and physical harm~\cite{button2017}. Targets often suffer intense emotional distress, including anxiety, depression, shame, and post-traumatic stress disorder (PTSD) symptoms~\cite{whitty2016online, acierno2019mental, coluccia2020online}, which can lead to erosion of trust ~\cite{hall2023trust} and social withdrawal ~\cite{gurun2018, robb2023, herrera2025}. The emotional toll is often felt more intensely than the financial harm~\cite{modic2015, whitty2016online}, can intensify existing health issues~\cite{button2014not}, and may result in suicidal ideation~\cite{cross2014challenges}. Consequently, effective interventions should be trauma-informed \cite{eggleston2025} and address scams' emotional dimensions. Absent from these scam frameworks is a perspective of the needs of targets across scam types, as scams unfold.

\subsection{How and Why Targets Engage With Scams}

We argue that scam susceptibility is a dynamic state influenced by targets' contextual risk factors~\cite{warford2022} and scammer tactics~\cite{hanoch2021}. Lack of digital literacy~\cite{downs2006, mcguinness2023, dhamija2006, oak2025reddit}, financial literacy~\cite{yu2021, zhang2022},  and scam literacy~\cite{oak2025reddit, bouma2025ccs} all affect susceptibility.  Timing~\cite{lyu2025, wright2023, butavicius2022}, environmental contexts~\cite{greene2018, wright2023}, and misplaced trust in platform security features~\cite{oak2025reddit} further shape target engagement. Scammers exploit both universal cognitive biases, such as optimism bias and sunk cost fallacy~\cite{downs2006, bouma-sims2025}, and specific situational risk factors, such as financial need, life transition, or loneliness~\cite{oak2025pig, button2014, stajano2011}. 

Social science research has analyzed psychological factors shaping scam susceptibility~\cite{button2014, buchanan2013, judges2017, norris2019, hanoch2021, deliema2020, cazanis2025}, using surveys to find correlations between susceptibility and traits like impulsivity~\cite{holtfreter2008}, high trust in others~\cite{fischer2013}, or neuroticism~\cite{tjondro2025}. Concurrently, computing literature has outlined the multi-stage mechanics of specific schemes like advance-fee fraud~\cite{ebot2023} and ``pig butchering” scams~\cite{oak2025pig} by detailing classic manipulation techniques like appeal to authority~\cite{milgram1963}, foot-in-the-door~\cite{freedman1966}, norm activation~\cite{schwartz1977}, and manufactured urgency~\cite{ariely2001}. Scammers apply these techniques by impersonating authority figures, escalating demands, invoking social obligations, and leveraging emotional manipulation to undermine rational decision-making~\cite{dove2021, langenderfer2001, stajano2011}. 

While researchers importantly suggest that scam awareness, digital literacy, and emotional support interventions are needed~\cite{oak2025reddit,bouma2025ccs}, existing work does not generally identify when, from a target's perspective, specific interventions would be effective. As Hassoun et al.~\cite{hassoun2023} argue, deficit-based information literacy interventions often occlude the factors that most motivate people to engage online and must be informed by contextual understandings of their specific motivations and practices to affect change. Chen et al. \cite{chen2025} found, for example, that increased engagement with scam prevention communication did not affect migrant workers' ability to assess scam risks or decrease their victim-blaming beliefs---but a reduction in fear and increased sense of self-efficacy (i.e., confidence that they could respond effectively to a scam, even if they perceived its risk as severe) did improve scam prevention skills. 

Building on these insights, we systematically explore how emotional motivations and scammer tactics drive target engagement with specific scams. Drawing on Roy et al.'s~\cite{roy2025} scammer manipulation tactics, we describe the emotional motivations that ``hook'' targets to engage in scams. We mobilize Warford et al.’s~\cite{warford2022} At-Risk Framework to understand what makes specific users more at risk of being targeted, hooked, and/or harmed by scams. We propose an expansion of the User States Framework~\cite{matthews2025} to map users' help-seeking needs to their emotional motivations and scam stage, identifying critical moments for interventions that respond to people's in situ needs.

\subsection{Online Help Seeking For Scams}

Formal scam reporting (\eg to government agencies, support organizations, law enforcement or platforms) is one path available to scam targets seeking help. However, scam incidents are chronically under-reported~\cite{aarp2025, oak2025reddit}. Studies highlight both personal motivations for justice and altruistic reasons for reporting~\cite{cross2018, meikle2024action}, but the majority of studies reveal persistent obstacles to formal help seeking. Research finds the most prominent obstacles are emotional, including shame, loss of trust, and self-blame~\cite{whitty2016, cross2015, modic2015, amirkhani2024, gurun2018, robb2023}. Victim-blaming norms and systemic issues like fragmented reporting channels and trauma-insensitive procedures isolate targets and dissuade them from reporting~\cite{anderson2021, cross2016improving, cazanis2025}.

While these barriers to formal help seeking persist, the emotional safety and pseudonymity of online forums can reduce stigma and foster candid peer support~\cite{choudhury2014, suler2004}. Within the computing literature, a small but growing body of work has examined scam-related help seeking in these spaces. Studies have categorized the types of help sought for image-based sexual abuse, including sextortion~\cite{wei2024}, documented how cryptocurrency communities educate users about scams~\cite{childs2024}, and identified common user knowledge gaps and peer support mechanisms in scam-focused subreddits~\cite{oak2025reddit, bouma-sims2025}.

While prior work characterizes the types of help-seeking interactions that occur online, a systematic understanding of how a target’s needs evolve throughout the course of a scam is still needed. In particular, many studies examine scams after they are over, with an emphasis on prevention and recovery support. Building on recent work examining the role Reddit communities can play in helping people who experience scams~\cite{oak2025reddit, bouma2025ccs}, we contribute a detailed synthesis of help needs throughout scams, including a nuanced look at targets' experiences 
as scams unfold. We draw on the User States Framework~\cite{matthews2025} to map a target's in situ emotions and help needs onto their stage in a digital-safety (\ie scam) event. This work contributes novel insights that can inform contextually-specific scam interventions. 

\section{Methodology}

\begin{figure*}[t]
    \centering
    \includegraphics[width=\linewidth]{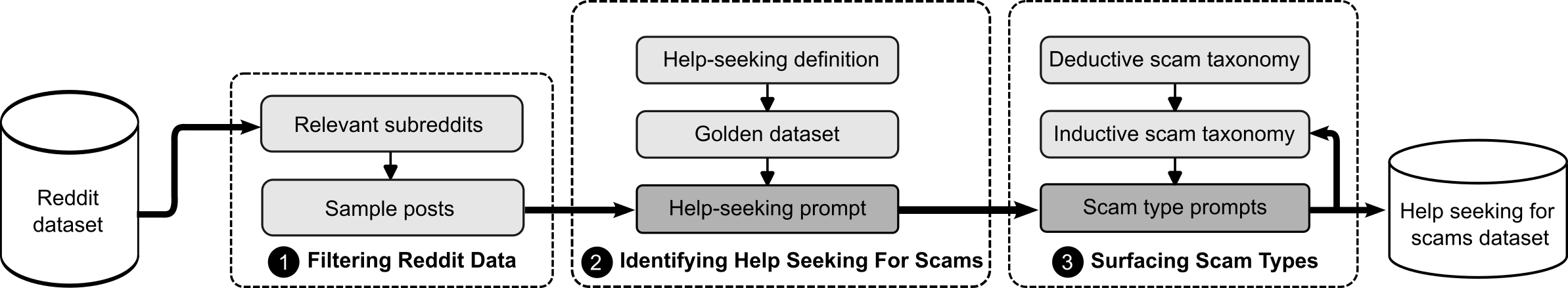}
    \caption{Overview of our data curation pipeline. Steps involving an LLM are in dark shaded boxes. Steps informing LLM prompt creation are in the lighter shaded boxes above these darker shaded boxes. \ding{182} We began by sampling an initial dataset of \dataASize posts from relevant subreddits. \ding{183} We examined each of these posts using an LLM to identify \dataBSize posts likely related to scam help seeking. \ding{184} We used manual review and LLM-assisted annotation to categorize posts into select scam types, sampling 500 posts across these scam types. We reduced this help-seeking dataset to a final set of 405 high-quality posts about 12 scam types after applying relevance and quality criteria in the qualitative analysis stage.
    }
    \label{fig:data-pipeline}
    \Description[Three step pipeline that filters a Reddit dataset into the help seeking for scams dataset.]{Figure 1 presents an overview of the data pipeline used to construct the help-seeking dataset from a pre-exisiting Reddit dataset. Reddit posts are filtered by subreddit and date, then identified as help seeking using LLM prompts informed by a definition and golden dataset. Help-seeking posts are classified into scam types using a scam type LLM prompt informed by deductive and inductive scam taxonomies, resulting in the final help-seeking dataset.}
\end{figure*}

We derive our analysis of help seeking for scams from data publicly shared on Reddit. We focus on Reddit due to its large user base, diverse communities, and the relative anonymity it affords, which may encourage users to seek help~\cite{choudhury2014, suler2004}. We adopt Wei \etal's mixed-methods approach---previously developed to examine help seeking for image-based sexual abuse on Reddit~\cite{wei2024}---to curate a dataset for in-depth qualitative analysis. Our goal in curating this dataset was to include a diverse range of known and emerging scams, not to seek a representative sample that covers all possible scams. We describe our approach to building a dataset for analysis, our coding process, ethical considerations, and limitations.

\subsection{Defining Help Seeking for Scams}
\label{subsec:help-seek-def}
To be considered for inclusion, a post had to involve help seeking for scams. We considered a post to involve help seeking for scams if the poster sought information, advice, or support to cope with a scam-related issue for themselves or someone else. We interpreted such support broadly to include emotional or therapeutic needs, which posters might express, for example, through venting, helping others, or telling their story about scams. We use Beals et al.'s definition of financial scams~\cite{beals2015}:

\pquote{Intentionally and knowingly deceiving the [target] by misrepresenting, concealing, or omitting facts about promised goods, services, or other benefits and consequences that are nonexistent, unnecessary, never intended to be provided, or deliberately distorted for the purpose of monetary gain.}{}

Like Beals et al.~\cite{beals2015}, we excluded identity theft, account compromise, or malware, focusing on events where targets were persuaded to actively engage (rather than unknowingly stolen from). We also excluded related attacks like phishing, account compromise, or malware, which scammers used as part of their set of tactics to carry out a scam.
\subsection{Curating a Reddit Dataset}

\paragraph{Filtering Reddit Data}
In our first step toward our goal to include a diverse range of known and emerging scam experiences in our dataset, we sought to filter Reddit data by sampling posts from relevant subreddits (Figure~\ref{fig:data-pipeline}, \ding{182}). To begin, we wanted to identify Reddit communities---called \emph{subreddits}---that were likely to have a high prevalence of help seeking for scams. To do so, we manually reviewed subreddits with more than 1,000 subscribers, accessing them via a site maintained by Reddit.\footnote{At the time of publication, Reddit lists its largest subreddits from most to least subscribers at this site: www.reddit.com/best/communities/1/.} 
We examined each subreddit, focusing on its description and recent posts, and selected communities that had recent posts and content relevant to scams.
We identified 134 subreddits with this method. These subreddits included those that were dedicated to platforms frequently targeted by scammers (e.g., \subreddit{r/cashapp}, \subreddit{r/facebookmarketplace}); that specialized in providing support related to scams (e.g., \subreddit{r/scams}, \subreddit{r/cryptoscams}); or that were about topics where scams were relevant, like digital safety or personal finances (\eg \subreddit{r/cybersecurity\_help}, \subreddit{r/povertyfinance}). 
Next, we sought to sample a diversity of scam-related data from these 134 subreddits. To do this, we randomly sampled up to 10,000\footnote{Setting a cap of 10,000 posts per subreddit provided a practical balance: it allowed for a broad coverage of active scam-related communities while still including smaller subreddits, supporting our goal of including a diverse range of scam-related help-seeking experiences.} \emph{original posts}---that is, the first post in a discussion thread\footnote{We include original posts because they focused on help seeking, and exclude comment threads which tended to focus more on help giving. For simplicity, we hereafter refer to original posts as \emph{posts}.}---from each of the 134 subreddits. We sampled these from an existing crawl of Reddit that adheres to robots.txt and crawler guidelines limiting to posts created between December 1, 2023 - December 1, 2024. Due to variations in community activity, not all subreddits had 10,000 posts within our analysis window. This resulted in a corpus containing \dataASize posts overall (with an average of 4,758 per subreddit).

\paragraph{Identifying Scam Help-Seeking Posts}
To identify help-seeking posts for scams in this corpus of \dataASize posts, we utilized a large language model (LLM) (Figure~\ref{fig:data-pipeline}, \ding{183}). We first constructed a golden dataset of 698 examples, balanced with 50\% positive examples (help seeking for scams) and 50\% negative examples (\ie not help seeking and/or not about scams). We created this golden dataset by randomly sampling 100,000 posts from the same 134 subreddits as the ``Filtering Reddit Data'' step, then used a keyword search (\eg ``scam,'' ``fraud'') to create a pool of candidate posts, and finally manually reviewed each post to select positive and negative examples of help seeking for scams. 
The golden dataset provided a ground-truth reference, allowing us to objectively iterate on prompts and measure how well the LLM identified help-seeking posts for scams. Our final classification prompt (Figure~\ref{fig:classification_prompt}) achieved a recall of 98\% and a precision of 68\% on Gemini 1.5 Flash. We optimized for recall over precision to support our goal of including a diversity of scam types. In total, our model identified \dataBSize potential help-seeking posts for scams. We used manual validation later in the process (described below) to ensure that posts in our final dataset met our definition of help seeking for scams.

\begin{figure}[t]
\centering

\begin{minipage}{\columnwidth}

   \begin{tcolorbox}[colback=customtaupe!25!white,
                    colframe=customtaupe!100!white,
                    arc=2mm,
                    ]
     \small {You are a financial fraud detection specialist. Your task is to analyze the provided Reddit post to determine if the author is either seeking help because they are a victim of financial fraud, a scam, or financial exploitation, or if the user is offering help or advice related to these topics.\\

\textbf{Focus specifically on posts where users are discussing fraud, scams, and financial exploitation.}\\  This includes terms like ``fraud,'' ``scam,'' ``identity theft,'' ``phishing,'' ``money laundering,'' ``investment fraud,'' ``financial abuse,'' ``embezzlement,'' and similar terms.\\

\textbf{Ignore posts that are:}\\
- Requests for customer support related to cryptocurrency transfer issues or lost cryptocurrency coins.\\
- Account recovery requests.\\
- Questions about estate planning or money disbursement.\\

\textbf{Subreddit}: \{\{ subreddit \}\}\\
\textbf{Reddit post}: \{\{ text \}\}\\

\textbf{Task 1:} Does the post meet the definition of seeking or giving help related to scams and fraud? Answer Yes or No.\\
\textbf{Task 2:} Provide a concise reason for why a post is, or is not, seeking or giving help for scams and fraud.

}
  \end{tcolorbox}
\end{minipage}
\caption{Prompt used to identify posts likely seeking help for scams. To favor recall over precision, the definition provided to the LLM goes beyond our definition of help seeking for scams. We relied on manual validation to remove imprecise or out of scope results.}
\label{fig:classification_prompt}
\Description[The LLM prompt in a blue box.]{Figure 2 presents the LLM prompt we used to identify help-seeking posts. It is blue border around native LaTeX text delineating the prompt from the rest of the paper.}
\end{figure}

\paragraph{Selecting Posts for a Diversity of Scam Types}
Next, we performed a series of manual and LLM-assisted annotation passes to select 500 posts for in-depth analysis (Figure~\ref{fig:data-pipeline}, \ding{184}), reflecting the number of posts we needed to reach meaning sufficiency given our research questions~\cite{braun2022}. To do so, we began with Beals et al.'s~\cite{beals2015} taxonomy to include a diversity of known scams, and supplemented it by examining a subset of randomly sampled posts from the data to include posts describing emergent scam types not included in their taxonomy (\eg cartel scams). Next, we used Gemini 1.5 Flash to annotate posts according to this initial set of scam types, and then drew a stratified random sample of 200 posts per scam type (2,000 total posts) from the 13,800 potential scam help-seeking posts. A single researcher examined the 2,000 posts to further refine scam types and confirm the posts were instances of scam-related help seeking suitable for in-depth qualitative analysis (\ie detailed enough to determine (a) what kind of scam it likely was and (b) what kind of help the poster was seeking). The researcher then met with the rest of the research team to discuss post inclusion. We performed this iterative examination twice more to refine our selections. 
To make inclusion decisions during this process, we used our judgment, knowledge of scams, and experience reading many help-seeking posts (for this and other work). To exclude posts from scammers themselves~\cite{Siu2023}, we especially scrutinized posts that mentioned specific individuals or services claiming to assist with scam recovery, or referenced potential scammer contact information.
Finally, from the remaining posts we randomly selected 50 per scam type to produce the 500 posts. This set of scam types was not exhaustive of all possible scams.

\subsection{Qualitative Data Analysis}
\label{sec:method-qual-code}
We used codebook thematic analysis (TA) to analyze the 500 posts, beginning with a deductive foundation from established frameworks and refining inductively based on emergent patterns in the data \cite{fereday2006}. We used codebook TA primarily for its ability to enable teams of multiple researchers to combine deductive (theory-driven) and inductive (data-driven) analyses of complex datasets~\cite{fereday2006, roberts2019codebookTA}. Similar to Wei et al.~\cite{wei2024}, we used a five-stage codebook TA~\cite{roberts2019codebookTA}: (1) initial code sourcing, (2) initial code development, (3) codebook design, (4) codebook application and reliability, and (5) interpretation. 

\paragraph{Initial code sourcing}
As part of the literature review covered in Section \ref{sec:related}, we identified potential codes from existing frameworks relevant to our research questions. We drew from five frameworks to source an initial set of candidate deductive codes related to online harms~\cite{scheuerman2021}, user states~\cite{matthews2025}, scam types~\cite{beals2015}, help-seeking needs~\cite{wei2024}, and scammer manipulation tactics~\cite{roy2025}. These candidate deductive codes would ultimately form the basic structure of our codebook (summarized in Appendix \ref{sec:codebook}), which we updated and expanded inductively. 

\paragraph{Initial code development}
We examined data to refine deductive codes and develop initial inductive codes throughout our data curating process (Figure~\ref{fig:data-pipeline}, \ding{182} \ding{183} \ding{184}). Three researchers read thousands of posts during this process and met to discuss impressions of the data and develop an early set of deductive and inductive codes. These three researchers applied these initial codes combined with open coding to 100 posts each (300 posts total). They then followed an iterative process of coding more posts, discussing, and updating codes, to ensure our codes were shaped by the data.

\paragraph{Codebook design}
From our code development process, we designed a codebook with code labels, definitions, instructions on how to apply the codes, and examples. We revised our codebook through our iterative coding process---including into the next stage of applying the codebook. 
The final codebook is summarized in Appendix \ref{sec:codebook} and consisted of:
\begin{itemize}

    \item \textit{Scam characteristics.} Codes included scam type (informed by Beals et al. ~\cite{beals2015}), scammer tactics (informed by Roy et al.~\cite{roy2025}), and harm type and details (informed by Scheuerman et al.~\cite{scheuerman2021}). 
    \item \textit{Help-seeking characteristics.} We refined the help-seeking types from Wei et al.~\cite{wei2024} to focus on four different kinds of help posters indicated they needed: support in understanding something (Sensemaking), advice on actions to take (Guidance), emotional support (Therapeutic), or direct intervention from others (External Action). 
    \item \emph{User state characteristics.} We refined the user states from the User States Framework~\cite{matthews2025}, splitting the Active Event category into distinct Diagnostic, Mitigation, and Denial substates (described in Section \ref{sec:help-seeking}). We also noted posters' emotions, motivations, and practices tried.
    \item     \textit{Target characteristics.} Codes included target demographics, the target's relationship with the poster (when they were different people), and at-risk groups or risk factors involved, following definitions in Warford et al.~\cite{warford2022}. 
\end{itemize}

\paragraph{Codebook application and reliability}
Four coders iteratively applied the codebook to our dataset of 500 posts (Figure~\ref{fig:data-pipeline}, \ding{184}). For reliability, we used a consensus coding approach~\cite{cascio2019}. 
Three coders independently coded one-third of the dataset, while a fourth coder coded the entire dataset to improve consistency. During this process, all coders kept memos and iteratively discussed disagreements in weekly meetings and interim online chat. Based on discussions, one coder resolved clear or already-discussed disagreements and marked issues for further discussion. After all disagreements were resolved, two coders split the data and checked all posts to ensure codes were consistently applied (further discussing any issues that arose). In this way, the team iteratively reached agreement on codes across all posts. We sought intercoder consensus, a form of reliability in qualitative analysis~\cite{cascio2019}, to support consistency of judgment among coders~\cite{boyatzis1998transforming} for data that were nuanced and codes that were not mutually exclusive.

During this iterative analysis, as our understanding of scams grew, we removed or recategorized a subset of the 500 posts.\footnote{We removed posts with insufficient detail and posts that appeared to be written by scammers. We did not resample replacement posts when observations were removed; instead, we prioritized maintaining our original sampling frame to avoid introducing selection bias and to ensure consistency in our analytic approach.} This process yielded a final dataset of 405 posts across 12 scam types (Figure~\ref{fig:data-pipeline}'s \textit{Help seeking for scams dataset}).

\paragraph{Interpretation}
Once the data were coded, we iteratively reviewed the coded data and notes taken as part of the coding process, and discussed themes. For example, we collated data by scam type, user states, emotions, and motivations. Over multiple discussions, we developed our five emotional motivation themes (Fear, Guilt \& Goodness, Trust, Hope, and Belonging). Our development of themes around help needs through scam lifecycles was supported by examining and discussing relationships in the data between help seeking and user state codes.

\subsection{Ethics}
While Reddit data is publicly accessible, we recognize that the Reddit users whose posts we analyzed did not explicitly opt in to the study. Following established HCI guidelines \cite{bellini2021so, williams2017, reagle2022, wei2022anti, wei2024} for quoted posts included in this paper, one researcher rephrased these quotes and systematically reverse searched each rephrased quote to ensure they did not return identifiable information. A second researcher reviewed the reworded data to ensure posts' original tone and meaning were preserved. To mitigate re-identification risk, we have chosen not to release the dataset. Although our institution does not mandate IRB review, we adhered to similarly strict ethical standards.

\subsection{Limitations}

Our study is limited to English-language posts on Reddit. Scam experiences discussed on other platforms or in other languages may differ. While data on Reddit language demographics is limited, a recent study found that over 97\% of Reddit content was posted in English~\cite{hoffa2021}. This indicates that the geographic distribution of Redditors posting in English likely resembles the overall distribution of active users. 
In Q4 2023, Reddit's Daily Active Unique (DAUq) user base was composed of 49.8\% US (36.4 million) and 50.2\% Rest of World (ROW) (36.7 million) users~\cite{reddit2024}. 

Our dataset reflected this diversity, containing references to numerous countries and regions---including the UK, India, Malaysia, Europe, and the US---though most posts did not indicate a specific region. The scam types in our study have been observed internationally. For example, Employment scams are a pervasive threat globally, with prior work demonstrating their high prevalence in countries like the UK~\cite{Ofcom2023} and India~\cite{Hirect2022, FCRF2024}. Charity scams are also a recognized global crisis, with documented cases targeting local and diaspora communities in countries like Malaysia~\cite{Bernama2024a,Bernama2024b}.

While the geographic composition of Reddit's active user base is roughly split between US and ROW users, we do not know the location of the specific posters that make up our dataset. We therefore cannot make claims about the impact of sociocultural differences on online help seeking from this data. Follow-up studies that critically analyze sociocultural differences in scam experiences would help explore whether our findings apply across and in specific contexts. We further discuss these potential adaptations in Section~\ref{sec:future-work}. 

As with other studies of scams \cite{oak2025reddit, bouma-sims2025, bouma2025ccs}, coverage of scams remains a key challenge and limitation of this work. In selecting scam-related posts for this study, we focused on including a diverse set of scam types and experiences rather than producing a representative sample of all scam-related help-seeking posts. These scam types included deductive and inductive types, as depicted in Figure~\ref{fig:data-pipeline}, \ding{184}, but did not include all possible scams (\eg sextortion~\cite{wei2024}).  

We used a single LLM for our analysis. Our recall analysis, which measured the model's ability to find all relevant posts using a manually verified reference, demonstrates that our LLM prompt identified 98\% of similar posts. To expand coverage of scams, we could add to the scam types we sought to find in Figure~\ref{fig:data-pipeline}, \ding{184}.

Because our analysis was focused on people's help-seeking behaviors and needs, we made inclusion and exclusion decisions based on self-reported accounts, which may contain technical inaccuracies, omit key details, or even be deceptive. Thus it is possible that posts were included that were not genuinely help seeking (\eg scammer posts) or about actual scams (\eg real debt collection or employment opportunities). Further, focusing on original posts misses help giving and interactions that might be present in comment threads---which have been explored in related work~\cite{oak2025reddit,bouma2025ccs} and could be further studied in future work. Finally, our scope was intentionally focused on financial scams based on definitions from prior work~\cite{titus2001, beals2015}, and thus our findings may not generalize to other forms of fraud, such as identity theft. 

\section{Results}

Our analysis revealed how scammers exploit targets' emotions to initiate and sustain engagement, and when and what kind of help posters seek as scams unfold. First, we introduce five main \emph{emotional motivations} scammers manipulated to engage targets.  Second, we show \textit{when} posters seek help (\eg before or after a target engages with the scammer) and \textit{what} they need at different critical moments of the scam, linking these needs back to the emotional context of the scam. Finally, we share how emotional motivations were intensified for at-risk users.

\begin{table*}
    \centering
    \renewcommand{\arraystretch}{1.2}

\begin{tabular}{p{1.5cm}|p{2cm}|p{9.2cm}|c}
\toprule
& & & \textbf{Posts} \\
\textbf{Main EM} & \textbf{Scam Type} & \textbf{Description} & \emph{(N = 405)} \\
\midrule

{\textbf{Fear} \newline \emph{N = 92}} & Cartel & Scammer impersonates a cartel or organized criminal group, threatening to harm a target unless they pay to resolve a fabricated offense. & 47 \\
\cline{2-4}
 & Phantom \newline Debt & Scammer impersonates a government agency or debt collector and intimidates a target into paying a fabricated or already paid debt. & 45 \\
\midrule

\textbf{Guilt \& \newline Goodness}  \newline \emph{N = 91} & Seller & Scammer deceives a target who is selling something online into providing the good without being paid (\eg using a payment method that is later reversed). & 31 \\
\cline{2-4}
 & Accidental \newline Payment & Scammer claims they inadvertently sent money to a target and requests the money be sent back. The original transaction is falsified or reversed. & 9 \\
\cline{2-4}
 & Charity* & Scammer impersonates a charity or otherwise seeks donations for a charitable cause from a target. & 51 \\
\midrule

{\textbf{Trust} \newline \emph{N = 69}} & Authoritative \newline Entity & Scammer impersonates a bank, service, platform, or public figure that a target nominally trusts to deceive the target into transferring funds. & 12 \\
\cline{2-4}
 & Romance* & Scammer romantically connects with a target online, manipulating the target’s trust and desire for companionship to extract money (often repeatedly and over time). 
 & 45 \\
\cline{2-4}
 & Personal \newline Relation & Scammer is or impersonates a family member or friend who deceives a target into transferring funds. & 12 \\
\midrule

{\textbf{Hope} \newline \emph{N = 153}} & Investment \& \newline Crypto \newline Culture* & Scammer deceives a target into transferring funds or crypto into a fake investment opportunity, typically promising large gains and belonging in a community of investors. & 43 \\
\cline{2-4}
 & Employment & Scammer offers a fake employment opportunity to deceive a target into transferring some of their own funds, sharing personally identifiable information (PII), working without pay, or acting as a money mule. & 56 \\
\cline{2-4}
 & Prize \& Grant & Scammer deceives the target into transferring funds to purportedly get access to a larger monetary payout (\eg a giveaway, a lawsuit payout).  & 30 \\
\cline{2-4}
 & Buyer & Scammer advertises a good on an online marketplace (that turns out to be fake or defective) that a target buys. & 24 \\
 \midrule
 \multicolumn{4}{p{15cm}}{*\textbf{Belonging} was also a main emotional motivation in certain scam types marked by an * above.} \\
\bottomrule
\end{tabular}

    \vspace{0.4em}
    \caption{Breakdown of the main emotional motivations (EM) manipulated by scammers, the different scams that hinge on these EMs, and the number of posts in our dataset.} 
    \label{tab:emotion_types}
    \Description[Categorizes twelve scam types by emotional motivation (fear, guilt/goodness, trust, and hope) and reports a brief description and post count for each category.]{Table 1 presents twelve scam types grouped by their emotionally charged motivations. For each type, the table provides a description and the number of posts that fall under that category. The emotional motivations used for grouping are fear, guilt and goodness, trust, and hope. Fear types include cartel and phantom debt collection. Guilt and goodness types include seller scams, accidental payment, and charity scams. Trust types include authoritative entity, romance, and personal relation scams. Hope types include investment and crypto culture, employment, prize & grant fraud, and buyer scams.}
\end{table*}

\subsection{Emotional Motivations Across Scams}
\label{sec:emotions}

Across the 12 scam types that posters mentioned (see Table~\ref{tab:emotion_types}), our analysis identified five \emph{emotional motivations} that scammers preyed upon in an attempt to overwhelm any initial skepticism on the part of a target and keep them engaged: (1) Fear, (2) Guilt \& Goodness, (3) Trust, (4) Hope, and (5) Belonging. As one poster emphasized, these emotions strongly influenced their actions despite being generally security-savvy:

\pquote{The invalidation from people, even close friends, has been incredibly difficult. I'm not an idiot, and I've even studied cybersecurity. The issue here wasn't technical though, it was emotional. I made a stupid decision in the moment due to my judgment being clouded by urgency and fear. Scams like this are more about manipulating your fear and isolation than about what you know.}{Phantom Debt scam}

{\noindent}We discuss each of these emotional motivations below, synthesizing tactics across individual scam types, since designing effective interventions likely requires tailoring warnings and guidance to each. We provide a high-level overview in Table~\ref{tab:emotion_types} of how four of these five emotions mapped to particular scam types as the \emph{main} emotional motivation. We share evidence that Belonging was a main emotional motivation in certain scam types and strongly influenced engagement across scam types (as presented in \ref{sec:belonging}), in line with prior work indicating that it is an important driver for online engagement generally~\cite{hassoun2023,hassoun2024mis}.

\subsubsection{Fear} Some scammers instigated and preyed upon fear, threatening targets with high-stakes consequences from powerful organizations or authorities. Examples included scammers who impersonated cartels, debt collectors, or government organizations (\eg HMRC, ICE).\footnote{HMRC is His Majesty's Revenue and Customs, the UK's tax, payments and customs authority; ICE is the United States' Immigration and Customs Enforcement.} These scammers threatened targets with severe physical violence (\eg murder), or escalating financial ruin (\eg credit score damage, garnished wages, lost housing) unless the target paid the scammer:

\pquote{A random number started sending me my home address and names and photos of my relatives, and is threatening to kill them. They've been harassing me all night to send the money. I'm scared it's a scam and I'll be left holding the bag but also scared about what happens if I keep saying no.}{Cartel scam}

\pquote{ This [medical company] is threatening my friend about a bill she allegedly owes. They are threatening to send the bill to debt collectors, even though we've sent so much evidence that she already paid the bill. It's been months of this. She's freaking out that this is going to mess up her credit score if it really does go to collection. What do we do?}{Phantom Debt scam}

Scammers flooded targets with repeated messages to urgently transfer funds to avoid any consequences. Even if targets were suspicious of the authenticity of the messages, they expressed being at a loss for how the scammer might have obtained so much personal information, making the target fear the risks of not engaging:

\pquote{I messaged a few escorts on [website]. I then got a message to my phone saying that I had wasted the girls' time and money, and that I needed to pay \$1500 or the organization has men tracking me and will come kill me. I've read posts here that seem similar but none that have this much info, like my cellphone and name and photo. I'm worried they'll find my address too. I just want to know if it is or isn't a scam. Please say I'm not alone in getting these messages -- I'm feeling paranoid and can't sleep.}{Cartel scam}

Scammers’ use of threats to impose urgency appeared to be a calculated attempt to overload and bypass the target's critical judgment. This tactic often weaponized existing anxieties about legal and financial authority, coercing immediate, irrational compliance. Scammers did not rely on any particular technical tactic to complete the scam transaction, seeming to trust that the emotional tactics eliciting Fear would compel some targets into complying.

Prior to seeking help, targets of Phantom Debt scams commonly discussed trying to verify the authenticity of a message, seeking proof from government institutions, credit bureaus, or their bank. However, many struggled with non-responsive customer service from these legitimate institutions. For Cartel scams, targets were often reluctant to involve law enforcement due to the perceived risk of disclosing how they came into contact with the scammer---which typically occurred after communicating with a sex work service.

\subsubsection{Guilt \& Goodness} Guilt \& Goodness-based scams presented targets with emotionally wrought or otherwise personalized scenarios where the target could purportedly help someone in need. For example, scammers posed as charities supporting people in conflict zones (a Charity scam), buyers in online marketplaces who needed a refund due to challenging life circumstances (a Seller scam), or someone who accidentally sent money that critically needed to be refunded (an Accidental Payment scam):

\pquote{I see so many [social media] stories about people in [conflict zone] needing help and money. Someone added me but they never show their face or really their surroundings either. I know it's probably a scam, but what if it isn't? What if no one else believes them and I'm all they've got?}{Charity scam}

\pquote{I was selling this item online and a guy said he'd give me \$90 via [Payment App] to reserve it for him to come get next week. He messaged today saying sorry but he didn't want it anymore. He said he knew he'd wasted some of my time and that he was fine with me sending back whatever I thought was fair. My plan was to give it back, but I just wanted to ask whether if I did there was any way for that money to get rescinded by [Payment App]? I don't wanna be scammed but I also don't want to be an asshole.}{Seller scam}

Scammers appeared to use the emotional tactic of norm activation~\cite{roy2025,schwartz1977,lea2009psychology}, which exploited the target's internal drive to uphold their moral identity as a ``good person,'' overriding skepticism and compelling the target to complete the transaction. Various transaction tactics were then used to defraud the target; commonly in our dataset, the scammer directed targets towards using a peer-to-peer (P2P) payment platform\footnote{P2P payment apps are digital platforms that allow users to send and receive money directly to/from one another. Examples include Zelle, PayPal, Venmo, and Cash App.} or making payments off-platform, effectively stripping away consumer protection and safety features of the original marketplace. In Seller and Accidental Payment scams, scammers often exploited a target's misunderstanding of complex financial systems and how funds might be processed or rescinded. Targets mentioned how funds from a payment appeared legitimately accessible in their accounts (which can occur when a scammer uses a stolen account or credit card to make the transaction). However, the scammer would request a ``refund'' (often via a different platform), after which the original transaction would be reversed (\eg by an anti-fraud system). 

Even if a target was suspicious of a scammer's request, they expressed a general concern they might be harming another person or that they could be someone's last line of support. To navigate this uncertainty, some targets reached out to their bank's fraud department or payment platforms for guidance. A few filed complaints with government agencies (e.g., the Internet Crime Complaint Center), seeking formal support. However, targets reported getting mixed guidance. For example, law enforcement and platforms instructed some targets to ``work it out with the person,'' ignoring the risk of a scam.

\subsubsection{Trust}
Trust-based scams exploited a target's existing trust, such as in a reputed business entity or public figure (Authoritative Entity scams); by establishing trust through forming a new romantic relationship with the target (Romance scams); or by leveraging an existing relationship with the target (Personal Relation scam): 

\pquote{My [mother-in-law] was trying to return something she bought online, and instead of signing into her account she looked up [retail company's] phone number and called it. Somehow the scammer got her to send \$11,000 and her social security number. I just don't get how she could do this. I'm angry but I also feel so sorry for her. She cried when I told her she wouldn't get her money back and that really broke my heart.}{Authoritative Entity scam}

\pquote{My mentally unwell uncle has been scammed out of his life savings by someone pretending to be a celebrity. It's been going on for over two years and he has literally nothing left. The scammer kept promising things like if he just gives another ten grand they can fly together on his private jet.}{Romance scam}

Many of the posters seeking help for Trust scams were family members of the target, in part due to the target's unwillingness to reconcile that their trust was being abused. Attempts to intervene often caused relational harm:

\pquote{We've all tried to tell them that it's a scam and that it's fake but they won't believe any of us because their feelings are so strong.}{Romance scam}

\pquote{When my sister brought it up, my aunt got angry and refused to talk to her for days.}{Romance scam}

Sometimes scammers were (or appeared to be) a personal relation of the target and abused this access to drain the target financially:\footnote{Some Personal Relation scams in our dataset described economic abuse, which is documented in literature on domestic and intimate partner abuse as one of many tactics abusers use to control survivors~\cite{duluth2025wheel}.} 

\pquote{My dad is being financially abused by multiple other family members. They control all his money and he is not even allowed his own bank account. I'm just trying to help him get out of there.}{Personal Relation scam}

\pquote{I got a request from my brother to send him \$35 on [Payment App] and then he would send me the same amount of money on [Different Payment App]. Is this a scam? I don't really get it.}{Personal Relation scam}

Trust-based scams sometimes achieved high severity and long duration by constructing a deceptive relational or institutional bond that superseded the target's critical judgment. Scammers’ primary tactic was impersonation, used to create a relationship of perceived credibility that drove targets to rationalize escalating financial demands. In Romance scams, the impersonation of a romantic interest appeared to exploit a deep-seated desire for companionship, causing targets to reject warnings from family members who attempted to intervene. In Authoritative Entity and Personal Relation scams, the scammer leveraged existing high-value trust—whether in an organization, public figure, family member, or friend—to circumvent skepticism and gain access to funds.

\subsubsection{Hope}
Hope-based scams presented targets with financial opportunities in the form of rapid investment gains,\footnote{While Trust scams may also present investment opportunities, their hook stemmed foremost from their trusted relationship with the target.} employment, prizes, or goods in online marketplaces:

\pquote{This [Messaging App] group that I joined had over 10k members and we were all earning huge money using this crypto wallet. When I tried to take my funds out they said I had to deposit some small percentage before taking money out for the first time. I paid it, like an idiot, but then I still couldn't withdraw anything and they said I needed to deposit more due to some technical complication with my wallet.} {Investment \& Crypto Culture scam}

\pquote{I'm new to the freelance life so I don't know much yet. I created an account on [Gig Work App], where I was contacted by a company. I communicated with them and ended up doing a transcription project for them. When I finished they told me I needed to pay \$150 in tax before getting the \$1600 for the job. Is this how it works or should I be worried?}{Employment scam}

\pquote{I got a phone call saying the lottery chose random phone numbers and I won millions of dollars and a new car. I don't know if it's legit but -- does this ever happen?}{Prize \& Grant scam}

Even if targets were skeptical, the potential financial upsides made them consider engaging:

\pquote{This person offered to help me pay off some medical bills. They said they'd transfer \$1200 but the transaction would cost \$30 so they'd need me to send them that upfront. This seems like a clear scam but I'm new to this and want to know for sure.}{Prize \& Grant scam}

These scams appeared to leverage the cognitive bias to prioritize a desired outcome—a job, an investment, or a prize—over objective risk assessment, causing targets to rationalize red flags. Exploiting Hope, scammers used the primary tactic of offering something of value to lure targets. Often adding urgency as a secondary tactic, the scammer created a high-stakes, time-sensitive environment that drove the target to complete the scam transaction through a technical action like sending money through a P2P payment app~\cite{roy2025,cialdini2008influence}. Scammers also convinced targets to accept the seemingly rational logic of paying a smaller upfront fee or withdrawal tax to receive a much larger promised reward~\cite{roy2025,freedman1966foot}, or set up fraudulent employment schemes where targets might unwittingly become money mules.\footnote{According to~\cite{fbi2025mules}, ``A money mule is someone who transfers or moves illegally acquired money on behalf of someone else.''}

\subsubsection{Belonging}
\label{sec:belonging}

Belonging was an emotional motivation which drove people to both engage with scammers and with other potential targets in their communities. People were drawn to scams by the need to Belong through personal companionship (especially common in Romance scams) and community membership (especially common in Investment \& Crypto Culture and Charity scams):

\pquote{My dad knew this much younger woman was probably scamming him the whole time. But even the small chance that the relationship was real kept him going -- he's just so lonely and keen to be with someone.}{Romance scam}

\pquote{A close friend is recently widowed from her (abusive) partner and seeking some purpose in her life. She joined a religious group that she's donating a lot of money to, and is also sending money to a group claiming that she's sponsoring a poor family overseas.}{Charity scam}

Uniquely, Crypto Culture targets often expressed low shame, as being scammed was a learning experience or rite of passage within the risk-tolerant community~\cite{Iamin2025, Olson2022,Cassino2023}. The act of (nonchalantly) sharing their experience to warn community members appeared to serve as a therapeutic function, supporting their need for Belonging: 

\pquote{The scammers are probably laughing at me, and that's fine, I messed up and I'm laughing at myself too. This post is to remind y'all (and me) that it's crazy out there.}{Crypto Culture scam}

These scams appeared to manipulate an emotional need for connection, which informational anti-scam advice alone might not resolve. For certain cryptocurrency investment groups, the aforementioned community norm of accepting risk might have contributed to people prioritizing social validation within the group over personal security~\cite{Iamin2025, Olson2022,Cassino2023}. In Romance scams, the desire for connection and purpose could be so strong that targets who had been defrauded continued to engage with scammers despite warnings from social relations. For some targets, the emotional utility of belonging to a community or being loved outweighed the financial trauma of being scammed, leaving the target susceptible to the next risky shared investment or relationship. The underlying emotional need to Belong could remain unresolved even after being scammed, leaving targets susceptible to repeat abuse. Scammers seemed to take advantage of the target's desire to Belong through tactics focused on in-group signaling~\cite{vanderDoes2022, Cassino2023} and love-bombing~\cite{cross2018}.

Across scam types, posters also informed others about particular scams or scammers, demonstrating a desire for Belonging through their concern for other community members. Many of these posts signaled Belonging to the community, using language and reflecting social norms common to that group. For example, this poster signaled their Belonging to the community of ``animal lovers'' in a post informing others about a particular scammer: 

\pquote{Please stay alert and don't support any campaigns run by these horrible thieves. They're actively stealing from animal lovers and the actual charities are suffering as a result.}{Charity Scam}
\subsection{Help-Seeking Needs Throughout Scams}
\label{sec:help-seeking}

Posters sought different kinds of help depending on \emph{when} during the scam they turned to Reddit. We describe posters' experiences in two dimensions: their emotional state at different points throughout the scam, and the specific types of help they sought in each state.

We use the states from the User States Framework\footnote{The states in the User States Framework are \textbf{Prevention} (when a person wants to minimize their exposure to future tech-facilitated attacks; they're likely to feel low or typical stress, unless they feel threatened), \textbf{Monitoring} (when a person wants to watch for signs of attacks; they're likely to feel low or typical stress, unless they feel threatened), \textbf{Active Event} (when a person wants to stop or otherwise respond to an attack(s); they're likely to feel high stress to panic), and \textbf{Recovery} (when a person wants to fix damage from the attack(s), determine what happened, and cope with trauma; they're likely to feel moderate to high stress)~\cite{matthews2025}.}~\cite{matthews2025} to map emotions and help needs to scam stages. This theoretical framework outlines people's likely emotions and needs before, during, and after technology-facilitated attacks. We expanded the framework and mapped these user states to scam stages that we identified as part of this analysis, as described in this section and depicted in Table \ref{tab:hs_us}. 

We identified four cross-cutting help needs throughout these user states: Sensemaking (when the poster wanted to understand something), Guidance (when the poster wanted to know what actions to do next), Therapeutic (when the poster wanted emotional support or expression), and External Action (when the poster asked others to take an action, such as reporting a scammer). We present the intersection of user states and help needs below (see also Table~\ref{tab:hs_us}). Table \ref{tab:helpseek_themes} summarizes our findings on common help needs throughout scams, indicating the user states when people especially needed certain types of help. 

\begin{table*}[t]
\begin{NiceTabular}{l|r|*{6}{c|}c|}
\cline{2-9}
 & \textbf{Scam Stage:} & Before & Hook & Some Harm & \multicolumn{2}{c|}{Ongoing Harm}  & After & \cellcolor{lightgray} \it Total \\ \cline{2-8}
 & \textbf{User State:} & Prevention &\multicolumn{3}{c|}{Active Event}  & Monitoring & Recovery & \cellcolor{lightgray} \it posts \\ \cline{4-6}
 &  &  & Diagnostics & Mitigation & Denial &  &  &  \cellcolor{lightgray} \\\noalign{\hrule height 1.5pt}
 \multirow{4}{*}{\begin{tabular}[c]{@{}l@{}}\textbf{Help-}\\ \textbf{Seeking}\\ \textbf{Type}\end{tabular}} 
 & Sensemaking & 12 & 71 & 136 & 13 & 7 & 23 & \cellcolor{lightgray} 251 \\ \cline{2-9}
 & Guidance & 9 & 13 & 80 & 24 & 15 & 17 & \cellcolor{lightgray} 136 \\ \cline{2-9}
 & Therapeutic & 65 & 7 & 47 & 8 & 11 & 72 & \cellcolor{lightgray} 149 \\ \cline{2-9}
 & External Action & 8 & 1 & 1 & 2 & 4 & 10 & \cellcolor{lightgray} 16 \\ \cline{2-9}
 & \cellcolor{lightgray} \it Total posts & \cellcolor{lightgray} 75 & \cellcolor{lightgray} 75 & \cellcolor{lightgray} 186 & \cellcolor{lightgray} 32 & \cellcolor{lightgray} 25 & \cellcolor{lightgray} 90 & \cellcolor{lightgray} 405 \\\noalign{\hrule height 1.5pt}
\end{NiceTabular}
\caption{The number of posts per user state and per help-seeking type, and how they map to scam stages we observed.}
\label{tab:hs_us}
\Description[Summarizes posts by help-seeking category and scam stage, mapping each stage to user states to show how experiences vary across the scam process.]{Table 2 offers a detailed breakdown of all posts in the dataset according to help-seeking categories shown in the rows, and scam stages shown in the columns. The scam stages are further mapped onto user states, providing an organized view of how posts distribute across help-seeking type and experiences at different scam stages.}
\end{table*}

\subsubsection{Prevention Before (and After) the Scam}

The Prevention state involved protective actions in the period of time \emph{before} a person encountered a scam. Prevention posts represented 19\% of the posts in our dataset. The vast majority of these posters warned others about a scam or scammer to help others avoid the scam before they encountered it, limit the damage scammers could cause, and/or signal Belonging to their community (\eg communities like a scam awareness group, crypto investors, online sellers/artists, etc.)---all Therapeutic purposes. For posters who did not indicate they had experienced the scam themselves, these posts seemed largely intended to \emph{help others}\footnote{We use italics to highlight help-need theme identifiers that are listed in Table \ref{tab:helpseek_themes}.} prevent being scammed.
For example:

\pquote{Basically there's this content creator and I've seen a ton of people sympathizing with him but I just need to warn everyone about this guy. [Describes scammer tactics.] I know it's a lot but it just felt important to share this to protect people. Stay safe out there!}{Charity scam}

Posters signaling Belonging to and concern for their community was also common, with phrases like \inlinequote{Be safe, stay alert folks,} \inlinequote{I hope sharing our experiences helps to protect others,} and \inlinequote{Y'all should be paid for your hard work and beautiful art.} Prevention with similar sentiments was also common \emph{after} scam encounters, as a part of Recovery (covered in Section \ref{sec:help-seeking-recovery}).

\subsubsection{During the Scam Active Event}

The vast majority of posters in our dataset requested help during an Active Event (72\% of posts), starting immediately after they became suspicious or aware of an active scam and extending through the time they were trying to stop or limit initial harm, often expressing emotions of worry or distress.\footnote{The definition of Active Event in Matthews et al. includes that ``users in this state are probably experiencing much higher than their normal amount of stress, even to the point of panic''~\cite{matthews2025}. While we could not directly assess stress, we considered expressions of emotion or distress during an event, together with other contextual clues in poster reports, when assessing the Active Event state.} We propose and describe three new substates below---Diagnostics, Mitigation, and Denial---to capture the nuanced help needs and emotions people expressed during an Active Event.

\paragraph{Diagnostics: Considering whether to engage}
People in the Diagnostics substate were considering engaging with a scam based on the scammer's \emph{hook}~\cite{roy2025} or manipulative tactics intended to draw them in (such as a scary threat or enticing prize), but had yet to do so or experience significant negative consequences. They were often in the process of identifying dubious behavior as a scam, expressing emotions of confusion and suspicion, but the emotional motivations luring them to the scam could be strong enough that they would consider engaging despite their suspicions.

Sensemaking was the most common type of help people sought during Diagnostics, especially requests for help \emph{identifying} whether something was a scam. In this example, the poster paused to consider a potential Employment scam, communicating the tension between their suspicions and hope of easing their financial need:

\pquote{Is this for sure a scam? I am desperate for work so I can't afford to miss out on an actual job opportunity, but I have that bad feeling in my stomach.}{Employment scam, Hope}

Posters also asked for \emph{explanations} of particular tactics, trying to make sense of an unknown scenario before engaging (e.g., asking how it might be harmful to send a buyer a refund). They also commonly wanted to know if others had \emph{experience} with the (yet to be confirmed) scam, seeming to use that experience as a proxy for understanding what might happen to them. When seeking Guidance in the Diagnostic substate, posters commonly asked \emph{whether to engage} with the (yet to be confirmed) scam, communicating their hesitation and thought process. For example, this poster asked for help \emph{whether to engage} (as well as \emph{identifying}) a potential Charity scam, motivated to consider it by a desire to help animals in need:

\pquote{Are these shelters in [country] a scam? I need advice on what to do, I'm worried that these poor animals are being starved. I don't know whether to help or not.}{Charity scam, Guilt \& Goodness}

Posters in Diagnostics had paused to seek help identifying scams before engaging or being harmed—this is a substate in which people would likely be primed to consider scam interventions, especially those that identify and explain scams and address the emotions motivating them to consider it (Hope, Fear, etc.). Such interventions in Diagnostics could be impactful in preventing harm (discussed further in Section 5).

\paragraph{Mitigation: Stopping further harm}
People in the Mitigation substate had experienced \emph{some harm} and wanted to prevent further harm. At this stage in the scam, they had been partially impacted (e.g., given some money, shared some personal information, or experienced significant emotional harm) and there was potential for more. Emotions spiked in this substate, including worry, fear, or even panic. Help-seeking needs focused on stopping or escaping the scam, understanding its potential impact, avoiding further harm, and soothing painful emotions.

Some targets skipped the Diagnostic substate by quickly being harmed by an exploitative scam tactic—such as scary threats for scams that preyed on their Fear, or fraudulent monetary transactions targets felt obligated to cope with as in Accidental Payment and Seller scams (detailed in Section \ref{sec:emotions}). These targets in Mitigation had the same Sensemaking needs as those in the Diagnostic substate—the most common still being requests for help \emph{identifying} scams. For example, a purchase transaction was underway before this poster/seller realized there may be a problem:

\pquote{I sold something to this guy and he sent me a check. I deposited it a few days ago but now he contacted me saying he had an unexpected issue and wants to return the item for a refund.}{Seller scam, Guilt \& Goodness}

Mitigation added new Sensemaking needs, especially for help \emph{judging} the situation (e.g., \inlinequote{should I be worried?} or \inlinequote{am I safe?}). With such requests, posters sought help from others to subjectively assess their situation. Posters also needed help \emph{predicting} the outcome of scams or related mediating actions. Common examples were posters who wanted to know if they might get their money back or what harm they might still face. For example, a poster asked for predictions about further harm from an Employment scam:  \inlinequote{I gave them a photo, my phone number, and email address. Am I just going to get a bunch of scam messages now?}

Beyond Sensemaking, posters in Mitigation had a notable need for Therapeutic help when coping with harm experienced---and potentially still pending---from a scam. They commonly expressed their emotions and sought \emph{reassurance} (e.g., \inlinequote{Can anyone put my mind at ease?} and \inlinequote{I can't sleep.}). These needs were often expressed in posts that were emotionally intense---for example, if the poster was scared or had lost a large sum of money.

Posters in Mitigation also commonly sought Guidance on how to exit the scam and mitigate harm. These inquiries asked \emph{how to} take an action (e.g., how to… respond, investigate, report, get my money back, cut ties), \emph{whether to} take an action (e.g., should I… respond, ignore, pay them, report to the platform, call the police, hire a lawyer), and open-ended requests for advice (e.g., what are my options? what else can I do?). For example, this poster experienced an Accidental Payment—a scam type that immediately put people in Mitigation after unexpectedly receiving a fraudulent payment—and asked a mix of \emph{open-ended} and \emph{whether to} questions:

\pquote{Someone sent me close to \$200, they asked me to send it back but I knew it was a scam and didn't respond. I didn't hear from them again but the money is still there a month later. Can I spend the money? Should I contact customer support? I don't know what to do.} {Accidental Payment scam, Guilt \& Goodness}

\begin{table*}[t]
\centering
\small
\begin{tabular}{p{3cm}| p{8cm} | p{2.9cm}}
\toprule
\textbf{Help-Need Type} & \textbf{Help-Need Themes} & \textbf{Prominent User States}\\ 
\midrule

Sensemaking
& \textbullet~~Identifying scams \textit{(Is this a scam?)} & Active Event\textgreater{}Diagnostics \\
\textit{Wanting to understand \newline something} &  \textbullet~~Explanations of scams or outcomes \textit{(How do they scam me if I give a refund?)} &  Active Event\textgreater{}Mitigation \\
& \textbullet~~Experiences with the scam \textit{(Anyone seen this?)} & \\
& \textbullet~~Judging the situation \textit{(Am I safe? Is this risky?)} & \\
& \textbullet~~Predicting the outcome \textit{(What might happen?)} & \\

\midrule

Guidance & \textbullet~~Whether to \textit{(Should I… respond? ignore? pay? report?)} & Active Event\textgreater{}Mitigation \\
\textit{Wanting to know what} & \textbullet~~How to \textit{(How do I… respond? investigate? report? get my money back?)} & Active Event\textgreater{}Denial  \\ 
\textit{actions to do next} & \textbullet~~Open-ended advice \textit{(What else can I do?)} & \\ 

\midrule

Therapeutic
& \textbullet~~Reassurance \textit{(Should I be worried?)} & Prevention \\ 
\textit{Wanting emotional support} 
& \textbullet~~Commiseration \textit{(Has this happened to anyone else?)} & Active Event\textgreater{}Mitigation \\ 
\textit{or expression} & \textbullet~~Venting \textit{(I just needed to share.)} &  Monitoring \\ 
& \textbullet~~Help others \textit{(Don’t fall for this scam. Stay safe.)} & Recovery \\ 

\midrule

External Action
& \textbullet~~Report a scammer \textit{(Help me report this scammer.)} & Recovery \\
\textit{Wanting help from a third} & \textbullet~~Share info about a scammer \textit{(Share this info to protect others.)} & Monitoring \\
\textit{party to perform an action} & \textbullet~~Vigilante behavior \textit{(Let’s get back at this scammer.)} & \\

\bottomrule
\end{tabular}
\vspace{0.4em}
\caption{Summary of our findings on posters' help needs throughout scams. For each help-need type, we provide an overview of common help-need themes and the prominent user states when posters especially needed certain types of help (though these are not exhaustive).}
\label{tab:helpseek_themes}
\Description[Outlines four help-seeking needs (sensemaking, guidance, therapeutic support, and external action) along with their recurring themes and associated user states.]{Table 3 provides an overview of the types of help users seek. It covers four main help needs: sensemaking, guidance, therapeutic support, and external action. For each of these needs, the table includes a list of recurring themes found within that help need type, as well as the user states associated with each type.}
\end{table*}

\paragraph{Denial: Harm continues unchecked} The Denial substate describes targets who did not acknowledge or believe that they were experiencing a scam, often leading to \emph{ongoing harm}. Denial is a psychological defense mechanism to protect against something that is threatening or emotionally painful~\cite{freud1937}. In our dataset, Denial was associated with targets who were motivated by a need for Belonging, especially companionship---many targets in Romance scams were in Denial (51\% of Romance scam posts). In these cases, the poster was typically a family member, friend, neighbor, or other relation of the target, who recognized or suspected the target was being scammed. 

Posters coping with a target in Denial often expressed a desire to convince the target they were being scammed or otherwise help them, especially by asking for Guidance on \emph{how to} convince them or that was \emph{open-ended}. For example, one poster wanted to convince their mother that she was in a Romance scam, and asked for advice:

\pquote{My mom is being scammed by someone pretending to be a famous actor. He's taken literally every penny she has. I tried showing her an article about other women falling for the same scam but I can't convince her that this guy isn't for real and I don't have the money to hire a lawyer or detective or anything. I don't know what to do and I'm so scared for her, would be grateful for any advice.}{Romance scam, Trust}

Posters sometimes sought Sensemaking (e.g., asking for \emph{explanations} or \emph{experiences}) or Therapeutic support by \emph{venting}: expressing emotions of frustration, worry, or hopelessness as they told stories of the target being scammed, often over extended timelines:

\pquote{I just don't know how to support her anymore after a year of this and 500k gone. To anyone here who went through this [romance scam] themselves: what helped you figure out what was happening? }{Romance scam}

\subsubsection{Monitoring Ongoing Scams}

The Monitoring state (6\% of posts) involved watching for signs of scams or scammer behavior, which typically co-occurred with Active Events when a scammer continued to cause harm. Most Monitoring posts were posters watching a target (who was in Denial) being harmed by the same scammer over time, with help needs and emotions overlapping with Denial. For example, to protect their parent who was in Denial for repeated Authoritative Entity scams, one poster reported they \emph{``have to manually block [scammers] every day''}, because if they did not, their parent \inlinequote{gives his bank details to anyone pretending to be famous or rich or offering him money - all my educational attempts have failed.} Some Monitoring posts overlapped with Prevention to \emph{help others} by informing a community about an active scammer, serving a Therapeutic need for the poster (as presented above for Prevention).  Monitoring was involved when the poster had been watching and compiling evidence on the scammer over time.

\subsubsection{Recovery After the Scam}
\label{sec:help-seeking-recovery} 

After doing what they could to mitigate the damage of a scam and stop it, and after signs of emotional distress had calmed, people entered the Recovery state (22\% of posts). In Recovery, posters' predominant concerns were to address lingering emotional harm, help others avoid the scam, and understand if anything they had lost was recoverable. They sought Therapeutic help for emotional harms, via \emph{reassurance} (to soothe still-strong emotions), \emph{commiseration} (to know they weren't alone), and \emph{venting} (to share their story and feelings).

As part of Recovery, posters commonly told their story of encountering a scam to \emph{help others} avoid it. Investment and Charity scam posts had a large portion that \emph{helped others}, showing their Belonging in the community of risk-tolerant investors or those who care about a charitable cause. For example: 

\pquote{I can't believe I was thinking about using my platform to help this awful person who is stealing money by using pictures from a real fundraiser for a family whose kid has cancer. I wanted to share to get the real family some money and spread the word about this scammer because they seem to have a lot of followers.}{Charity scam}

Another related form of help seeking was asking for External Action, which posters often did during Recovery. Common External Action requests included asking others to \emph{report} a scammer, \emph{share} information about a scammer, and engage in \emph{vigilante behavior} (confirming results from Oak and Shafiq~\cite{oak2025reddit}).

\subsection{Factors that Elevate Risk }
\label{sec:at-risk}

Contextual risk factors are factors known to augment or amplify the chance that someone will experience technology-facilitated attacks or disproportionate harm from such an attack~\cite{warford2022}. We identified four related factors---financial insecurity, legal precarity, gig work \& precarious labor, and neurodiversity \& mental health---that influenced who experienced scams, their likelihood of experiencing harm at specific scam stages, and the types of help they needed.\footnote{Our description of risk factors is by no means exhaustive; we aim to illustrate why contextual risk factors should be considered as part of designing effective interventions.}

\subsubsection{Financial Insecurity}
Previous research has found that people experiencing financial and housing insecurity are vulnerable to scams~\cite{sleeper2019tough}. In our data, this insecurity led targets to engage with scams, skipping the Diagnostic substate of Active Event out of intense Hope and desperation for a potential job or basic necessity like food and housing, regardless of the perceived risk involved: 

\pquote{Someone from HR reached out and told me to send details to some random email address. It didn't feel right but the money was pretty good and I needed a job so desperately.}{Employment scam, Hope}

\pquote{My family have been really struggling with money but they found a [social media] page advertising free stuff. I was so excited to help so I entered all our details (email, phone number, birthday, etc.) but it was a scam and now my family is upset.}{Prize \& Grant scam, Hope}

This insecurity also increased the impact of financial and emotional harm:

\pquote{The car he sold me stopped working almost immediately. He knew that money was all I had left, and that I'm homeless and was gonna live in this car. When I called to try to trade the car back for my money he told me his friends would come kill me.}{Buyer scam, Hope}

People who had prior experiences with debt had a hard time distinguishing Phantom Debt scams from potentially legitimate requests, requiring extra layers of verification compared to other potential targets: \inlinequote{I'm getting calls about some debt that I paid ages ago. Am I being scammed? What if I can't find proof that I paid it back then?}

Financially insecure targets often struggled to find the requisite time and mental space to diagnose and cope with scams (experiencing the ``resource or time constrained'' risk factor identified by Warford et al.~\cite{warford2022}). Sleeper et al's~\cite{sleeper2019tough} work highlights several circumstances that directly elevate the risk of harm for financially insecure individuals like those in our study: These individuals were vulnerable to engaging with suspicious low-cost online options (such as low-cost housing scams) out of urgent necessity, even when they suspected the deals were ``too good to be true.'' Further, the need to use personal information actively for essential services (like applying for government benefits) meant they were habituated to providing sensitive data (like their social security number) to websites, which increased their overall risk of scams or phishing. When harm did occur, the impact was disproportionately severe; even small financial losses could affect basic necessities like food or rent, and limited resources made it difficult to recover from issues like identity theft, as they lacked the funds or standing with financial institutions to repair damaged credit.

\subsubsection{Legal Precarity}
Legal precarity---either due to immigration status, previous incarceration, or court judgments---also affected how people navigated scams, and could amplify Fear manipulated by some scammers. Prior work has suggested that migrant communities may be more vulnerable to scams, in part due to language barriers and regularly having to share sensitive personal information with authorities~\cite{simko2018computer}. This was reflected in our data:

\pquote{My elderly parents got a letter saying they owe thousands from some medical procedures my dad got years ago. They don't speak English well enough to understand the letter or contest it themselves, but my mom thinks it's something to do with their insurance and my dad wants to just not pay. My reaction was to tell them to request a letter with the alleged charges rather than sending any money over the phone.}{Phantom Debt scam, Fear}

Similar to financial insecurity, prior legal charges made it difficult to distinguish scams from potentially legitimate judgments. Formerly incarcerated individuals faced significant employment barriers, increasing their financial insecurity, and risked experiencing elevated harm if they unknowingly participated in illegal activities like money muling. Many hesitated to seek formal help and were quicker to respond to scammers out of fear:

\pquote{I'm trying to help someone who is struggling to deal with fines from court charges that they thought they settled over a decade ago. They got a letter that looks like a legal summons but it said just their name and the state - there was no listed judge and it didn't say what the summons was for. This person responded right away by calling the number on the letter and money was taken out of their bank account days later. I need help finding information and figuring out the laws here.}{Phantom Debt scam, Fear}

Scammers impersonating legal organizations manipulated the perceived power differential between government entities and formerly incarcerated individuals to scare targets into engaging with scams. Language barriers also affected targets' ability to diagnose scams and contest scammer-inflicted issues. These barriers and risks have also been observed for undocumented immigrants~\cite{guberek2018}, illustrating how ``legal or political'' risk factors like immigration status and ``underserved accessibility needs'' like language barriers can affect scam susceptibility and impacts~\cite{warford2022}.

\subsubsection{Gig Work and Precarious Labor} 
Gig workers\footnote{Gig workers are workers who perform ``task-based work conducted outside of a formal employment relationship, often paid per task and to various degrees governed by digital platforms''~\cite{katta2024, woodcock2019}.} face precarious labor conditions, including lack of legal protection and uncertain access to regular income~\cite{katta2024, barwulor2021sexwork}. People applying for minimum wage jobs are more likely to be targeted by scams~\cite{simko2018computer}. In our data, posters seeking general or informal labor found scams hard to distinguish from legitimate opportunities, increasing their risk of engaging with the scam. Those who primarily relied upon online sales for income risked experiencing more harm from scammers (in comparison to someone selling a one-off product online for supplementary income). Artists and sex workers who created online content were targeted due to their reliance on public-facing digital platforms and because the precarious nature of their labor made institutional legal or financial recourse riskier:

\pquote{I'm just starting out as an adult content creator, and using social media to get followers. I'm really behind on bills. I get a lot of messages about being paid 3k a month if I just pay them a fee first. These are a scam, right?}{Employment scam}

When scams made these targets lose trust in online platforms, it impacted their ability to find new jobs~\cite{ravenelle2022} or engage in online marketplaces and communities:

\pquote{After losing money because of a guy and now being harassed by him, I'm afraid to take custom requests because it could be him again, or some other scammer, and I don't want to put in all that work and money for nothing.}{Seller scam}

These online content creators were ``prominently'' visible to scammers and therefore susceptible to being targeted~\cite{warford2022}. Sex workers encountered platform-enforced and legal barriers to safe employment practices~\cite{barwulor2021sexwork}.  Many precarious labor workers across employment types expressed being ``resource or time constrained''~\cite{warford2022}, which meant scams could profoundly harm their ability to earn income~\cite{warford2022}. 

\subsubsection{Neurodiversity and Mental Health}
Neurodiversity and mental health conditions affected how people reacted to scams and assessed risk. Targets who reported related conditions generally expressed intensified emotional harms (especially from scams that preyed on Fear or in a Mitigation state during an Active Event) and needed more Therapeutic support, even if they knew they were likely safe:

\pquote{I almost fell for the scam and I'm so scared that all my devices have viruses. I'm having panic attacks and my OCD means these thoughts won't leave my head even though I know they're irrational. Please don't say anything scary here since I'll just freak out more.}{Employment scam}

Some such targets expressed that they impulsively engaged with scams before thinking through the potential consequences:

\pquote{After looking at the posts and comments here I decided to send some messages to the scammer to mess with them a bit, but now I'm worried they might be able to find me. I have ADHD so can be a little reckless, I also have quite severe anxiety. Would appreciate any help.}{Cartel scam}

Prior traumatic experiences also exacerbated the harm targets experienced:

\pquote{The scammer has my address and sent me some really upsetting photos but also a video. I didn't see any other posts here mention a video so are you sure this is a scam? I have trauma from some stuff I saw as a kid so the photos that the scammer sent were really triggering, literally shaking right now.}{Cartel scam}

Mental health struggles stemming from abuse and isolation sometimes led targets to engage with scams: 

\pquote{My abusive husband of 30 years died and I was in a horrible place and all alone. I needed a car so I found someone at a dealership to sell me one. He took advantage of my situation and stole tens of thousands of dollars from me for ``investments.'' He lost his job at the dealership and vanished.} {Romance scam}

Such posters expressed both elevated anxiety about scammers attacking them and fears that they could not confidently protect themselves or recover from a scam's emotional effects, demonstrating ``underserved accessibility needs''~\cite{warford2022} which could increase their risk of experiencing harm.
\section{Discussion}

In this section, we discuss cross-cutting themes from our study and how our findings could inform the design of interventions that use the \emph{right message}, at the \emph{right time}, to address the \emph{right need} for people targeted by scams---building on Intille's~\cite{Intille2004} concept of ``just-in-time'' messaging. Scam interventions are likely to be more effective when they deliver needed information using emotionally-sensitive messaging at key moments. For example, an intervention could diffuse the emotions scammers are attempting to manipulate (\ie the \emph{right message}), explaining a relevant common scammer tactic to give the person diagnostic support (\ie the \emph{right need}). The intervention could display this warning when a person encounters a potential scam online (such as a job seeking site or on social media) (\ie the \emph{right time}).
We propose potential support strategies and indicate when and what kinds of interventions might be particularly useful.

\subsection{Right Message: Emotional Motivations}

Since emotions influence how information is received and processed~\cite{langenderfer2001, wright2023}, we argue that scam interventions may be more effective if they deliver emotionally-sensitive responses to address the emotional components driving engagement. We identified five emotional motivations that scammers manipulated to hook and engage targets: Fear, Guilt \& Goodness, Trust, Hope, and Belonging. Messaging in scam interventions should consider these emotional motivations when attempting to convince targets to avoid or stop engaging the scammer. Without addressing the underlying needs driving people's engagement with scams, interventions---particularly those focused on scam literacy, education, and other informational messaging---may not be effective. In particular, Belonging was a strong motivator for scam types in our data that involved Denial and extensive ongoing harm. It is unlikely that people motivated by Belonging would be responsive to primarily informational messaging, as we found with targets of Romance scams. Investment \& Crypto Culture scam targets appeared to be partially driven by a desire to socially belong in a community that valued risk-taking and amused indifference to large losses~\cite{Cassino2023, Iamin2025, Olson2022}, also suggesting that primarily informational messaging may not be effective.

Our findings support prior work on misinformation susceptibility that argues for addressing people's underlying motivations for engagement, which are often emotional and social~\cite{Phadke2025}. This prior work finds that both youth~\cite{hassoun2023} and adults~\cite{hassoun2023, kim2024,ren2023} share misinformation to meet social and emotional needs. Like the scammers in our research, misinformation creators explicitly target social and emotional needs to motivate people to engage~\cite{hassoun2024mis}. Prior research has found that people also assess information and make credibility judgments together as members of communities~\cite{geeng2020, hassoun2023}.

\subsection{Right Time \& Need: Help Seeking for Scams}

Our findings detail people's help-seeking needs at key moments before, during, and after scams. Below, we suggest how these findings can be used to tailor interventions to help needs at different scam stages.  

\paragraph{Before a scam} 
Posts related to the Prevention state demonstrated Reddit community members' interest in intervening to help others avoid scams and the important Therapeutic needs of Belonging and altruism this type of sharing supported. This community of helpers could be supported by their inclusion in scam prevention interventions (\eg through reporting suspected scams to others or platforms). 

\paragraph{Hook} 
Targets considering a scammer's emotionally manipulative hook in the Diagnostics substate (during an Active Event) needed help identifying and avoiding harm from the scam. In this critical pre-harm window, targets needed help to address the emotions motivating them to engage (\eg Hope, Fear, etc.) and contextually-specific Sensemaking support to explain the scam mechanics. 

\paragraph{Some harm} 
During a scam and after experiencing some harm in the Mitigation substate of Active Event, targets needed Sensemaking help to understand what was happening to them, Guidance on preventing further harm, and Therapeutic support to cope with escalated emotions like worry and fear. 

\paragraph{Ongoing harm} 
Targets in the Denial substate of Active Event often experienced ongoing harm from scams and were not receptive to help. Family and relations described Monitoring these ongoing scams while feeling frustrated, worried, or hopeless, and expressed a need for Guidance on how to help the target mitigate harm and move toward Recovery; such help could be vital in stopping the severe harm often involved in these cases of repeated financial exploitation by scammers. 

\paragraph{After escaping a scam} 
Posters in the Recovery state primarily needed Therapeutic support to cope with the emotional impact of the scam and to connect with others (such as by reporting the scam to help others avoid being hooked).

As outlined above, the three new Active Event substates we introduced for scams---Diagnostics, Mitigation, and Denial---enabled us to provide more granular break-down of a target's needs and emotions at key moments when they were facing an active scam.

\subsection{Potential Interventions}
Here, we bring together these findings to propose potential interventions that address the right need, at the right time, with the right message for scam targets.\footnote{Please note that the design of effective digital-safety interventions is very challenging to get ``right'' in practice---something prior work has argued is a wicked problem~\cite{matthews2025}. Determining how to design scam interventions in privacy preserving, safe, and ethical ways that are effective on different platforms and for people world-wide, is an area for future work involving multi-disciplinary experts and communities. Our work can help inform starting points for evidence-based intervention explorations, but the ideas presented here should not be considered prescriptive or comprehensive.} We acknowledge that while our dataset captures a global user base, the predominance of English-language posts means that the help-seeking behaviors observed could reflect Anglophone cultural norms (e.g., specific expectations of institutional utility). Consequently, the interventions proposed below should not be viewed as universal solutions, but rather as strategies to be adapted to local sociocultural contexts (particularly regarding institutional trust, shame/stigma, and community support). 
We discuss these adaptations further in Section \ref{sec:future-work}.

\paragraph{Fear Interventions}
Fear-based scams often involved scary threats and were largely reported in our dataset as being carried out in direct messaging. Interventions could focus on mitigating fear with emotional reassurance and informing targets of red flags that suggest what they are experiencing could be (or even likely is) a scam that poses no real physical or financial danger. This sensemaking support could help demystify the threat and empower targets to disengage. Interventions could also give people clear guidance on how to verify potential scammer claims, like how to certify debt collection notices. Such interventions are likely most beneficial in the Diagnostics and Mitigation substates, when targets are actively receiving threats likely to elevate their distress.

\paragraph{Guilt \& Goodness Interventions}
For scams that manipulate a target's sense of Guilt \& Goodness, interventions could help targets identify how scammers manipulate their desire to help others, explaining the emotional and technical tactics as they are encountering them. For example, within messaging platforms, especially those on P2P marketplace sites and apps, an intervention could detect conversational red flags,\footnote{The privacy implications of such detection would need to be carefully considered.} such as a someone using emotionally manipulative language to compel an off-platform ``refund,'' a common tactic in Seller scams. A just-in-time alert could intervene to discourage the platform switch, focusing on immediate information to explain how the scammer's tactic works both emotionally and technically. Such guidance may be most effective in the Diagnostic state, as people attempt to determine a solicitation's authenticity.

\paragraph{Hope Interventions}
Targets of hope-based scam types could be dealing with financial insecurity, unemployment, and inexperience. Interventions could provide guidance sensitive to these vulnerabilities without discouraging legitimate opportunities. In the Diagnostic state, for example, just-in-time warnings about requests for upfront fees or personal data on job-seeking platforms could help prevent job-seekers from engaging with Employment scams. Many job-seeking posters in our study expressed both suspicion and hope (or even desperation) when asking about potential employers, suggesting that interventions could confirm red flags in job postings and support people in seeking further information to verify the legitimacy of the organization. Adaptations may be necessary for contexts where and people for whom the reliability of formal institutions (e.g., law enforcement, government agencies, banks) is low \cite{Hadler2025, Yeager2017}. In such circumstances, interventions might be more effective if they encourage sensemaking and verification through community-based organizations rather than institutional authorities.

In our data, most targets of these scam types reached Mitigation or Recovery, which suggests the value of interventions at these stages in places where targets seek help. This could include guidance on what to do if the target has already shared personal information with the scammer (in the Mitigation state), or support for navigating the legal or financial consequences of unwittingly participating in illegal activities (\eg as a money mule) in the Recovery state. 

\paragraph{Trust Interventions}
For scams that rely on impersonation tactics, like Authoritative Entity scams, encouraging the target to pause and guiding them in verifying the identity of the scammer is critical. An emotionally-sensitive just-in-time intervention could specifically target the moment a scammer is attempting to exploit a user's faith in an organization or person. Many banking platforms have versions of these just-in-time messages, intervening at the moment of transfer. P2P platforms, which posters commonly mentioned scammers manipulating, could potentially benefit from similar messaging. In lieu of the platform itself implementing such messaging, a browser or app (with user consent) could provide this Diagnostic support.

While Romance scams are among the most widely researched in the HCI literature~\cite{amirkhani2024, carter2021, cross2023, whitty2016} and targets engaging with them are deeply emotionally motivated~\cite{cross2018}, our research suggests they may be the least amenable to just-in-time interventions. The Denial state prevalent in Romance scams presents a unique challenge, as the targets themselves were largely not receptive to interventions and may have been socially isolated by scammers, making it important to direct Guidance and Therapeutic support towards their concerned family and friends. Our data indicated that targets of these scams had sometimes experienced repeat victimization, which might be because the underlying emotional motivations that initially led to engagement may have remained unresolved.

\paragraph{Belonging Interventions}
While technical and educational interventions may help mitigate immediate financial harm, longer-term recovery from emotional harm often requires Therapeutic support. As discussed, while an emotionally-motivated desire for Belonging made some scams---like Romance, Investment \& Crypto Culture, and Charity scams---seem more compelling to targets, the emotional need for community connection also appeared to motivate targets recovering from scams to offer Therapeutic support to others across scam types. This expression of Belonging in the form of help-giving served an important Therapeutic function for targets seeking Recovery, while providing educational help to those in Prevention. 

Previous research~\cite{oak2025reddit, bouma2025ccs} already demonstrates the importance of online communities like Reddit in providing peer-to-peer therapeutic support and reassurance for targets. Platforms could contribute to this ecosystem by improving formal scam reporting mechanisms (a form of External Action). Enhancing these systems could potentially allow people, as part of their Recovery process, to anonymously tell their story through a publicly visible, scam-specific platform reporting function—a Therapeutic help-giving activity that could both aid their recovery and help warn others. Other research has found that people learn behavior-impacting lessons from digital security stories told by others, especially those which elicit an emotional response~\cite{Rader2012}, suggesting that helping people tell their scam stories to others may serve a valuable preventative function.  In sociocultural contexts where financial loss carries especially deep stigma or shame \cite{Fan2024}, interventions may need to prioritize anonymous, private, or one-way support channels rather than public forum participation.

\subsection{Future Work}
\label{sec:future-work}

We propose that future work could investigate how our findings might apply to other types of scams not covered in our study, such as gambling scams, or beyond scams to other types of technology-facilitated attacks. We also discuss below how our work could inform future studies on automatic detection, user research, and the effects of sociocultural variation on help-seeking norms and interventions. 

\paragraph{Automatic detection and user research}
Automatic detection of potential scams could improve the timing and customization of just-in-time messages. Future work could examine whether our characterizations of scams by emotional motivations and user states could help inform such automatic scam detection. Any scam detection efforts would need to carefully account for user privacy and safety. In lieu of detection, designers could focus intervention efforts on functions that scammers misuse to manipulate targets, such as P2P sales, P2P payments, direct messaging, and more.

Translating this study's findings to the design of effective digital-safety interventions in practice will require extensive outreach efforts and rigorous applied research, as noted above. For example, researchers experimented for more than four years on just-in-time messaging and visual design for malware, phishing, and SSL warnings in Chrome to improve effectiveness~\cite{Reederetal2018, Feltetal2014, AkhaweFelt2013, Feltetal2015, Almuhimedietal2014, Feltetal2016}. As a baseline, future work could compare and test user reactions to emotionally-sensitive anti-scam messaging versus purely informational ones to maximize protective efficacy across different scam stages. Likewise, future work is needed to iteratively design and evaluate just-in-time scam messaging to establish and improve effectiveness on specific platforms.

\paragraph{Sociocultural variation}
Our findings and their implications for interventions are likely affected by sociocultural help-seeking norms, suggesting areas for future work to understand help seeking for scams in different geographies. For example, online help-seeking models based on data from the United States and the Philippines failed to generalize to students in Costa Rica, who engaged in significantly more offline, collaborative help seeking~\cite{Ogan2014}. A study with Chinese first-generation college students found they preferred ``one-way'' rather than ``interactive'' online help seeking~\cite{Fan2024}. Another study with ethnic minority first-generation college students found they underutilized support services due to fear of relational damage~\cite{Chang2020}. These differences may affect online help seeking in ways that require further study. Some findings may generalize across contexts: anonymous support, for example, has been shown to reduce stigma and foster candid peer support in multiple cultural contexts~\cite{Xu2022, Pretorius2019, Prescott2017}.

The therapeutic value of public processing of trauma online likely varies based on cultural context, as evidenced by a study comparing Chinese Australian and European Australian trauma survivors~\cite{Jobson2024} which found that cultural beliefs were associated with different PTSD symptom outcomes. This suggests that a trauma-informed approach \cite{eggleston2025} to scam story sharing~\cite{Rader2012} requires ``cultural humility''~\cite{Ranjbar2020}, as it is essential to first understand how potential recipients ``understand and prefer to communicate about traumatic stress and the process of healing''~\cite{Ford2015}.

Scam experiences and help-seeking behaviors are also influenced by socioculturally-mediated trust in institutions~\cite{Hadler2025}. For many communities, institutions like law enforcement and banks have not historically been trustworthy~\cite{Yeager2017}. Many Trust scams preyed on targets' trust in institutions, and falling prey to such scams decreased that institutional trust for some targets. Fear scams also preyed on targets who had prior negative experiences with institutions, like debt, incarceration, or legal issues. Such interactions between at-risk user status~\cite{matthews2025} and institutional trust~\cite{Hadler2025} as they relate to scam susceptibility and help-seeking behavior merit further study, particularly toward improved intervention support.
\section{Conclusion}
This research, grounded in a user-centered analysis of posts on Reddit that sought help for scams, details how emotional motivations and scammer tactics interact to shape a target's susceptibility. We mapped help-seeking timelines and needs as scams unfold, highlighting when and why people reach out for support. We detailed implications for intervention design, including guidance on how user experience insights can inform timely, effective and empathetic scam prevention strategies.

Our findings indicate that scammers adeptly leverage targets' emotional motivations (including Fear, Hope, Guilt \& Goodness, Trust, and Belonging), using both emotional and technical tactics. Targets predominantly seek help during the ``Active Event'' state—when they are actively trying to stop or mitigate immediate harm—though ``Recovery'' is also a significant phase for seeking support. The three new Active Event substates we introduced for scams—Diagnostics, Mitigation, and Denial—elucidate targets' different needs at key moments during an active scam. These help-seeking needs include Sensemaking (to understand the situation), Guidance (for next steps), Therapeutic (for emotional support), and External Action (for third-party intervention), underscoring the need for tailored support.
 
Notably, emotions were an important factor in scams---as motivators for engaging scams, part of the harm experienced, and dictating the help needed. Our findings further demonstrate that factors such as financial insecurity, legal precarity, platform dependent labor, neurodiversity, or mental health issues can heighten a target's risk and lead to more severe harms. These insights could inform the design of contextually sensitive interventions that move beyond technical scam detection, providing holistic support that addresses the real-life sensemaking, guidance, therapeutic, and external action needs of scam targets at specific critical stages of scams.

\begin{acks}
We would like to extend a very big thank you to the many people who have supported this work, including Amanda Walker, Bryant Gipson, Luca Invernizzi, reviewers of this paper, and more. 
\end{acks}

\bibliographystyle{ACM-Reference-Format}
\bibliography{references_clean}
\appendix

\section{Codebook}
\label{sec:codebook}

\subsection{Scam characteristics}

\textbf{Scam type} (informed by Beals et al. ~\cite{beals2015}): Select the code from below that best captures the type of scam (see Table \ref{tab:emotion_types} for descriptions of each scam type):
    \begin{itemize}
        \item Cartel
        \item Phantom Debt
        \item Seller
        \item Accidental Payment
        \item Charity
        \item Authoritative Entity
        \item Romance
        \item Personal Relation
        \item Investment \& Crypto Culture
        \item Employment
        \item Prize \& Grant
        \item Buyer
    \end{itemize}

{\noindent}\textbf{Scam details} (informed by Roy et al.~\cite{roy2025}): Notes on tactics and mechanisms of the scammer, focusing on how they interact with the target to induce emotions, impel action, manipulate, deceive, and/or defraud them. Examples:

    \begin{itemize}
        \item Appeal to emotion
        \item Norm activation
        \item Make a threat
        \item Impersonate a person
        \item Impersonate an org/authority
        \item Offer something of value
        \item Urgency (to induce a concern, such as fear)
        \item Urgency (to indicate scarcity)
    \end{itemize}

{\noindent}\textbf{Harm} (informed by Scheuerman et al.~\cite{scheuerman2021}): Select all codes from below that describe the type of harm self-reported by the poster, or take notes if they do not apply:

    \begin{itemize}
        \item Financial: ``Financial harm is defined as material or financial loss, including the loss of digital assets like accounts.''~\cite{scheuerman2021}
        \item Emotional: ``Emotional harm ranges from an annoyance (at its least severe) to a stressful or traumatic emotional response (at its most severe), whether fleeting or long-lasting.''~\cite{scheuerman2021}
        \item Relational: ``Relational harm is defined as damage to one’s reputation or their interpersonal, professional, or larger community relationships.''~\cite{scheuerman2021}\footnote{The fourth harm type from Scheuerman et al.~\cite{scheuerman2021}---Physical harm---was not present in our data.}
        \item Legal: The poster reports harm related to legal issues.
        \item PII lost: The poster reports that the target lost personal information to the scammer.
        \item Time lost: The poster explicitly reports time lost to the scam as a hardship. (Do not interpret the time spent on the scam as a harm unless it is explicitly described as a hardship.)
        \item Stolen labor: The poster reports the scam involved performing uncompensated labor.
        \item None: The poster has not reported that the target has been harmed. This captures cases where the target catches the scam before the point of harm. 
    \end{itemize}
    
{\noindent}\textbf{Temporality}: For cases when harm has occurred, take notes on how long the target was involved with the scam or any other interesting time-relevant information, if explicitly described. 

{\noindent}\textbf{Repeat victimization}: Indicate ``Yes'' and optionally take notes when a target has been victimized by more than one scam.

{\noindent}\textbf{Additional notes}: Notes on harm details, such as what was lost, monetary amounts, quotes describing emotional harm, etc.

\subsection{Help-seeking characteristics}

\noindent\textbf{Help needs} (informed by Wei et al.~\cite{wei2024}): Indicate the type of help the poster describes needing. Select any codes below that apply:

\begin{itemize}
    \item Sensemaking: Expresses a need for help in trying to understand something related to a scam.
    \item Guidance: Expresses a need for help about what actions to take related to a scam.
    \item Therapeutic: Expresses a need for emotional support, such as reassurance, validation, self expression, etc. Includes cases when the poster has a strongly emotional tone, indicating a need to vent. Includes cases when the poster is helping others.
    \item External Action: Expresses a need for a third party to take direct action to help the poster/target. 
\end{itemize}

\noindent\textbf{Help giving:} Notes on help given to others, such as informing others of a scammer and their tactics, explaining scams, sharing their story of a scam encounter, etc.

\subsection{User state characteristics}

\noindent\textbf{States from User States Framework~\cite{matthews2025}}: Indicate the states during which the poster is seeking help. Select any state code below that applies. If Active Event is selected, also select only one Active Event substate code that best captures the substate:
\begin{itemize}
    \item Prevention: When a person wants to minimize their exposure to future tech-facilitated attacks (or help someone else do so); they’re likely to feel low or typical stress, unless they feel threatened. This tends to be done when a person is not actively coping with an attack. 
    \item Active Event: When a person wants to stop or otherwise respond to an attack(s). This tends to include the period of time starting immediately after a person starts suspecting or becomes aware of the attack and extends through the time they are trying to stop or limit initial harm, often expressing worry or distress. If a person is unable to understand or stop an attack, or if attacks overlap, this could be an extended period of time.
    \begin{itemize}
        \item Diagnostics: When a person is considering responding to a suspected scam, but has yet to do so or experience significant negative consequences. This tends to be when a person is in the process of trying to identify whether suspect behavior is a scam; they may feel confusion or suspicion, but also may feel motivated to engage. This also includes cases when the poster does not yet realize they are potentially experiencing a scam, but they worry something is wrong before engaging and a researcher can identify a likely scam from their description.
        \item Mitigation: When a person has been partially impacted by the scam and there is potential for more harm. Emotions spike in this substate, including worry, fear, or even panic. A person likely wants to stop the scam, understand its potential impact, avoid further harm, and soothe emotions.
        \item Denial: When a person targeted by a scam does not acknowledge or believe there is any fraudulent activity in their interactions with a scammer. Another person suspects or recognizes the Active Event and wants to help the target. The target may not seem worried or distressed, though others may express worry or distress due to the target's Active Event.
    \end{itemize}
    \item Monitoring: When a person wants to watch for signs of attacks; they’re likely to feel low or typical stress, unless they feel threatened. This can be done separately from an event or attack, but also in response to threats or attacks. Monitoring can commonly overlap with Active Event or Recovery—after a user is attacked, a typical response is to vigilantly monitor for more to occur.
    \item Recovery: When a person wants to fix damage from the attack(s), determine what happened, and cope with trauma; they’re likely to feel moderate to high stress. This tends to start after the acute stress or panic of the event has subsided (though people in this state may still express ongoing mental health challenges). This state can extend over a long time period, since resolving damage and coping with trauma from an attack can be lengthy processes.
\end{itemize}

\noindent\textbf{Emotion notes}: Notes on what emotions the poster is expressing.

\noindent\textbf{Motivation notes}: Notes on what motivations the poster is expressing.

\noindent\textbf{Practices tried}: Notes on previously attempted practices or remediation steps mentioned by the poster.

\subsection{Target characteristics}

\noindent\textbf{Demographics}: Notes about the demographics of the target and poster (if different).

\noindent\textbf{Target's relationship to the poster}: Notes about the relationship between the poster and the target, if they are different people.

\noindent\textbf{At-risk user}: Notes about whether the target and poster (if different) mention an at-risk group or contextual risk factor in relation to the scam (following definitions in Warford et al.~\cite{warford2022}).
\end{document}